\documentclass[aps,prb,preprint,superscriptaddress,12pt,a4paper]{revtex4}

\usepackage[usenames]{color}
\usepackage{amsmath,amsfonts}
\usepackage[centerlast,sc,small]{subfigure}
\usepackage{graphicx}
\usepackage{graphics}
\usepackage{epsfig}
\usepackage[justification=justified]{caption}
\usepackage{pifont}
\usepackage{epstopdf}

\def\Redstar {{Re_{\delta^*}}}
\def\Retheta {Re_\theta}
\def\dstar {{\delta^*}}
\def\utau {{u_{\tau}}}

\def\uinf {{U_e}}
\def\Up {U^+}
\def\yp {y^+}
\def\dbl {{\delta}}
\def\dblo {{\delta_0}}
\def\lz {\lambda_z}
\def\lzp {{\lambda_z}^+}
\def\meter {\rm{m}}
\def\milimeter {\rm{mm}}
\def\ms {\rm{m.s^{-1}}}
\def\degree {\rm{\ensuremath{^\circ}}C}
\def\amplstreaks {\widehat{A}_{st}}

\begin{document}
% \doi{10.1080/14685240YYxxxxxxx}
%  \issn{1468-5248}
%  \jvol{00} \jnum{00} \jyear{2010}
% 
% \markboth{Pujals, Cossu \& Depardon}{Journal of Turbulence}

% \articletype{GUIDE}

\title{Forcing large-scale coherent streaks in a zero pressure gradient turbulent boundary layer}

 \author{Gregory Pujals}
 \affiliation{LadHyX, CNRS-\'Ecole Polytechnique, F-91128 Palaiseau, France}
 \affiliation{PSA Peugeot Citro\"en, Centre Technique de Velizy, 2 Route de Gisy, 78943 V\'elizy-Villacoublay Cedex, France}

 \author{Carlo Cossu}
 \affiliation{IMFT - CNRS, All\'ee du professeur Camille Soula, 31400 Toulouse, France}

 \author{Sebastien Depardon}
 \affiliation{PSA Peugeot Citro\"en, Centre Technique de Velizy, 2 Route de Gisy, 78943 V\'elizy-Villacoublay Cedex, France}

% \author{Gregory Pujals$^{\rm a,b}$$^{\ast}$\thanks{$^\ast$Corresponding author. Email: gregory@ladhyx.polytechnique.fr
% \vspace{6pt}}, Carlo Cossu$^{\rm a}$$^\dagger$\thanks{$^\dagger$ Present address: IMFT - CNRS, All\'ee du professeur Camille Soula, 31400 Toulouse, France
% \vspace{6pt}} and Sebastien Depardon$^{\rm b}$\\\vspace{6pt}  $^{\rm a}${\em{LadHyX, CNRS-\'Ecole Polytechnique, F-91128 Palaiseau, France}}; $^{\rm b}${\em{PSA Peugeot Citro\"en, Centre Technique de Velizy, 2 Route de Gisy, 78943 V\'elizy-Villacoublay Cedex, France}}\\\vspace{6pt}%\received{\today}}
% \maketitle

\begin{abstract}
Large scale coherent streaks are artificially forced in a well developed turbulent boundary layer at $\Redstar \approx 1000$ using an array of cylindrical roughness elements. Measures of the velocity field with particle image velocimetry reveal the presence of well reproducible, streamwise oriented, steady coherent streaks.
We show that the amplitude of these coherent streaks transiently grows in space. The position of the maximum amplitude is proportional to the spanwise wavelength of the streaks and the most amplified spanwise wavelength is of very large scale $\lz \approx 6 \dblo$. These results are in good agreement with the recent predictions based on the optimal transient growth analysis of turbulent mean flows. 

% \begin{keywords}Turbulent boundary layer; Streaks; Optimal perturbations
% \end{keywords}\bigskip

\end{abstract}
\maketitle

%%%%%%%%%%%%%%%%%%%%%%%%%%%%%%%%%%%%%%%%%%%%%%%%%%%%%%%%%%%%%%%%%%%%%%%%%%%%%%%%%%%%%%%%%%%%%%%%%%
\section{Introduction}\label{sec:intro}

The presence of streaky structures is a commonly observed feature of wall-bounded turbulent flows. 
These structures range from the well documented near-wall streaks  populating the buffer layer with a mean spanwise spacing of $\approx 100$ wall-units \cite{Kline1967,Smith1983,Kim1987}, to the newly emphasized large and very large-scale structures of the logarithmic layer \cite[][]{Townsend1961,Kovasznay1970,delAlamo2003,Tomkins2003,Hutchins2007} whose spanwise spacing is proportional to the shear flow outer length scale $\dbl$. Those structures can contain as much as $40\sim60\%$ of the turbulent kinetic energy and are responsible for $40\%$ of the shear stress production \cite[][]{Balakumar2007}.  
If the self-sustained mechanism by which the near-wall structures reproduce themselves in the buffer layer begins to be well understood \cite{Jimenez1991,Hamilton1995,Waleffe1995}, this is not the case for the large-scale structures whose formation and sustain mechanisms are still the object of debate.

In laminar wall-bounded flows high-energy streaks can be induced by low-energy streamwise vortices via the lift-up effect  \cite[see e.g.][]{Moffatt1967,Landahl1980,Schmid2001} which leads to the algebraic growth of the streaks in the linear approximation. This growth, that is only transient in the presence of viscous effects \cite{Gustavsson1991}, is related to the non-normal nature of the linearized Navier-Stokes operator and can be very large when optimized \cite[][]{Butler1992,Trefethen1993,Schmid2001}.
An appealing idea is that the lift-up mechanism is active, in some average sense, also in completely developed turbulent flows.
Butler \& Farrell \cite{Butler1993} have computed the optimal perturbations supported by a turbulent channel flow. 
In \cite{Butler1993} the turbulent mean velocity profile of Reynolds \& Tiedermann \cite{Reynolds1967}, based on Cess' eddy viscosity model \cite[][]{Cess1958}, is used as base flow. However, the molecular viscosity is used in the linearized equations for the coherent perturbations. With this approach it was found that the most amplified disturbances are spanwise periodic and streamwise uniform vortices with an optimal spanwise wavelength $\lz \approx 3h$, $h$ being the channel half-width. Optimal spanwise wavelengths of $\lzp=100$ were retrieved by constraining the optimization time to appropriate values of order of the eddy turnover time \cite{Butler1993}.
Reynolds \& Hussain \cite{Reynolds1972} had however convincingly shown that small coherent perturbations are better described by linearized equations that use the eddy viscosity in equilibrium with the turbulent mean flow.
This formulation has been therefore adopted by del~Alamo \& Jim\'enez \cite{delAlamo2006} who have then recomputed the optimal perturbations and growth in this framework. They find two peaks of optimal energy growth without any restriction on the optimization time, which is also an output of the optimization algorithm. The primary peak scales on outer units (i.e. $h$) and is associated with an optimal wavelength $\lz=3h$. The secondary peak scales with inner (wall)-units and is associated with $\lzp=100$. 
In this study \cite{delAlamo2006} however, the energy amplification associated with the primary peak (outer units) was observed to decrease with the Reynolds number, contrary to what is observed in the laminar case. 
Actually, some inconsistency in the formulation used in \cite{delAlamo2006} was later found by Pujals et al. \cite{Pujals2009} who have therefore re-performed the analysis using the consistent linearized operator. With the revised formulation \cite{Pujals2009} it is found that the primary peak, associated with the large scales, is obtained for $\lz=4h$ and that the associated maximum energy growth increases with the Reynolds number. 
The secondary peak is associated with near-wall structures with $\lambda_z^+ \approx 100$ and is substantially unchanged with respect to the results of \cite{delAlamo2006}.  

The fact that a mean lift-up is also observed in the fully turbulent case and that the optimal amplification of large-scale streaks increases with the Reynolds number suggests that these structures could be artificially forced for control purposes. The forced streaks could be used for the passive control of turbulent flows and the efficiency of the control, given by the transient energy growth, would increase with the Reynolds number. The forcing of optimal streaks has actually been shown to be an effective way to passively control laminar boundary layers \cite[see e.g.][]{Cossu2002,Fransson2006}. 

The analysis of the turbulent Poiseuille flow of Ref.~\cite{Pujals2009} has therefore been extended \cite{Cossu2009} to the turbulent boundary layer on a flat plate with zero-pressure gradient using the mean velocity profile of Monkewitz et al. \cite{Monkewitz2007}. Also in this case two peaks of optimal energy growth are observed. The secondary peak scaling in near-wall units is very similar to the one found for the Poiseuille flow and is associated with structures that correspond well to the most probable streaks observed in the buffer layer. The energy growth associated with the primary peak, that increases with the Reynolds number, is however larger than the one observed for the Poiseuille flow. The outer peak optimals are obtained for very large wavelengths of order of $8\dbl$ for high Reynolds numbers and $6\dbl$ for low Reynolds numbers. The corresponding optimal vortices and streaks were found to be almost Reynolds-number independent when rescaled with the proper outer length scale (i.e. $h$ in the turbulent channel case and a modified Rotta-Clauser length in the turbulent boundary layer case \cite{Cossu2009}). 

The theoretical prediction of the optimal growth of coherent streaks is based on strongly simplifying assumptions and,
to our knowledge, no experimental observation of the transient growth of coherent streaks in a fully developed turbulent mean flow is currently available. This is especially a problem in the case of the turbulent boundary layer. Actually, in the turbulent plane Poiseuille flow, large-scale structures with the optimal spanwise periodicity of $4h$ are naturally observed even if they are not the most energetic ones (see \cite{delAlamo2003} and discussion in \cite{Pujals2009}). The same is observed in turbulent Couette flow where the $\lambda_z \approx 4-5h$ spanwise scale scales observed in unforced direct numerical simulations \cite[e.g.][]{Komminaho1996} match the theoretical predictions \cite[][]{Hwang2009} and have been experimentally forced  \cite{Kitoh2008}. 
%even if their transient amplification has not been explicitely measured.
Naturally occurring very large-scale structures with $\lz \geq 6\dbl$ have not been detected yet in turbulent boundary layers where the most energetic structures have a spanwise scale of rather $\lambda_z \approx \dbl$.
It could be that the larger optimal structures, even if largely potentially amplified by the mean lift-up, are not able to self-sustain in the boundary layer because they are not selected by the other processes involved in the self-sustained process \cite[see e.g.][]{Hamilton1995,Schoppa2002}. It could however also be that, after all, no mean lift-up exists and that other mechanisms are responsible for the existence of very large-scale streaks. 

The first scope of this study is to verify if the transient growth of coherent streaks can actually be experimentally observed in a turbulent flow. This is an important step aimed at validating or discarding the theoretical predictions of coherent transient growth in turbulent flows. As a second objective, we want to verify to which extent the most amplified measured scales match the predictions of the linear optimal perturbation analyses. 
In order to have reproducible results we have decided to artificially force large scale streaks in the turbulent boundary layer on a flat plate. We measure the induced velocity fields by particle image velocimetry. The coherent streaks develop from coherent vortices forced by an array of cylindrical roughness elements evenly spaced in the spanwise direction. The approach is very similar to the one already used to force moderate amplitude streaks in the laminar boundary layer \cite[][]{White2002,Fransson2004,Fransson2005,Fransson2006}. The experimental setup is described in section~\ref{sec:Setup}. 
The turbulent mean flow obtained in the absence of forced streaks is described in section~\ref{sec:BaseFlow}, while the measured coherent streaks are described in section~\ref{sec:Streaks}. We anticipate that we effectively observe the transient growth of the streaks and that the most amplified spanwise scales match well the theoretical predictions. These findings are further discussed in section~\ref{sec:Conclusions} where some conclusions are also drawn.

%%%%%%%%%%%%%%%%%%%%%%%%%%%%%%%%%%%%%%%%%%%%%%%%%%%%%%%%%%%%%%%%%%%%%%%%%%%%%%%%%%%%%%%%%%%%%%%%%%
\section{Experimental setup} \label{sec:Setup}

The experiments have been conducted in the PSA Peugeot Citr\"oen in-house facility in the Aerodynamics Department in V\'elizy. The wind-tunnel is of closed-return type with a $800 \milimeter$ long test section and a $0.3 \meter \times 0.3 \meter$ cross sectional area. The temperature can be kept constant and uniform within $\pm 0.5\degree$. The contraction ratio is $8$ and the velocity can be controlled from $7 \ms$ up to $45 \ms$.  The turbulent boundary layer forms on a flat plate placed on the roof of the test section. 
We denote by $x$, $y$ and $z$ the streamwise, wall-normal and spanwise directions respectively.
Transition to turbulence is tripped $50\milimeter$ downstream of the contraction using a strip of sandpaper and the turbulent velocity profile is well developed sufficiently downstream. 
%  placed above the test section (resulting in $(x,z)$ planes). A $28mm$ optical lens is used resulting in $300mm\times220 mm$ field of view. The laser sheet is $1mm$ thick and, in order to ensure the convergence of the mean velocity fields, $600$ pairs of images are acquired. All the data presented here are acquired at $Y=2mm$ from the wall. 

The velocity is measured using the Dantec Dynamics' Flow Manager particle image velocimetry (PIV) system associated with a double-cavity pulsed Nd:Yag laser ($120 \rm{mJ}$ per pulse) and a $1024\times1280$ Hisense Mk2 CCD camera with a $8$ bit dynamical range. The seeding particles are DEHS droplets which diameter is smaller than $1 \rm{\mu m}$.
The velocity fields in the presence of streaks are acquired in $x-z$ planes parallel to the wall. 
For this type of measures, the camera is placed below the test section and is equiped with a $28 \milimeter$ optical lens is used resulting in $300\milimeter \times 220\milimeter$ field of view. The streaks are measured on a plane situated at $Y=2\milimeter$ from the wall and the laser sheet is $1\milimeter$ thick. In order to ensure the statistical convergence of the mean velocity fields, $600$ pairs of images are acquired at a frequency of $4.5Hz$. 
A different setting, where the measures are taken on vertical $x-y$ planes, is used to measure the wall-normal velocity profiles in the absence of streaks. In this case the camera is equipped with a $105\milimeter$ optical lens. The field of view is $20\milimeter \times 14\milimeter$ and $1500$ pairs of images are acquired. 

All the acquired data are post-processed with Dantec Dynamics' Flowmap software. With both configurations (mean velocity measurements and streaks visualizations), an image processing is used to reduce the impact of optical disturbances: a reference image where each pixel is the mean pixel value evaluated over the complete set of recorded images is computed. Final images are obtained by subtracting this reference image to the raw images and then adding an offset in order to keep the pixel value within the $8$ bit range. Mean velocity profiles are obtained by a four-step adaptive correlation calculation (the final size of the interrogation areas being $16\times8$ pixels) with $8$ iterations and $50\% \times 75\%$ overlap in respectively the streamwise and normal directions. Velocity in the streaks are calculated by a two-step adaptive correlation method (with interrogation areas of $16 \times 16$ pixels) and $50\% \times 50\%$ overlap. In both cases, a peak validation filter set to $1.2$ is used to minimize the number of spurious vectors \cite[see][]{Adrian1991}.
As already mentioned, the streaks are forced by using one array of evenly spaced cylindrical roughness elements that introduce nearly-optimal streamwise vortices in the boundary layer, as already done in the laminar boundary layer \cite{White2002,Fransson2004}. The roughness elements are pasted on a very thin ($<0.1\milimeter$ thick) scotch tape and painted in black to avoid reflections due to the laser sheet.

% \newpage
%%%%%%%%%%%%%%%%%%%%%%%%%%%%%%%%%%%%%%%%%%%%%%%%%%%%%%%%%%%%%%%%%%%%%%%%%%%%%%%%%%%%%%%%%%%%%%%%%%
\section{The turbulent boundary layer base flow} \label{sec:BaseFlow}

\begin{figure}
 \begin{center}
 \resizebox*{8cm}{!}{\includegraphics{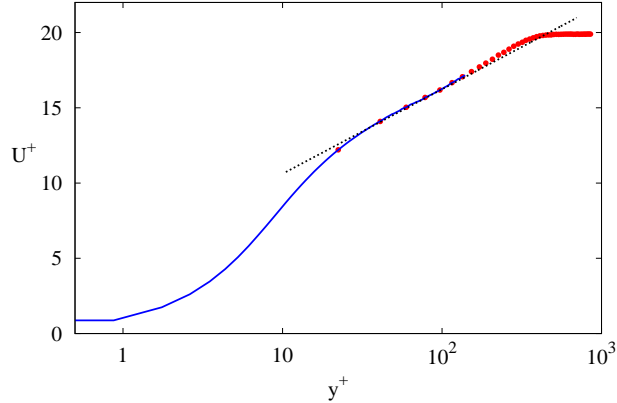}}
 \caption{Mean velocity profile $\Up$ plotted against $\yp$. {\textcolor{red}{$\bullet$}} Experimental data measured with PIV. {\textcolor{blue}{\boldmath{$\relbar\!\relbar\!\relbar$}}} best fit. {\textcolor{black}{\boldmath{$\cdots\,\cdots$}}} Analytical law \unboldmath{$\Up=\log(\yp)/0.41+5$}.}
%  \caption{Mean velocity profile $\Up$ plotted against $\yp$. {\textcolor{red}{$\bullet$}} Experimental data measured with PIV. {\textcolor{blue}{\boldmath{$\leftrightline\!\leftrightline\!\leftrightline$}}} best fit. {\textcolor{black}{\boldmath{$\cdots\,\cdots$}}} Analytical law \unboldmath{$\Up=\log(\yp)/0.41+5$}.}
 \label{fig:Velocity}
 \end{center}
\end{figure}

Preliminary experiments have been performed in order to characterize the baseline turbulent boundary layer obtained in the absence of forced streaks. The free-stream velocity is set to $\uinf=20\ms$ to obtain a well developed boundary layer in the short streamwise extent of the test section. The evolution of the free-stream velocity in the streamwise direction indicates a slight favourable pressure gradient. The mean velocity profiles are measured along the centerline of the test section. The spanwise homogeneity is checked measuring the spanwise variations of the dimensionless mean velocity $U(x,Y=2\milimeter,z)/\uinf$, at a distance of $2\milimeter$ from the wall. These variations are within a $1\%$ range across the entire width of the test section. At the given large speed $\uinf$, the boundary layer is very thin: at $x_0=110\milimeter$ the boundary layer thickness\footnote{In this study, we use the $\delta_{99}$ definition} is $\dblo=5.4\milimeter$ which, together with the few number of particles near the wall, makes difficult precise measurements of the velocity by PIV in the near-wall region. 
% \begin{table}
%  \tbl{Turbulent boundary layer parameters. $\beta_P=\dstar/\rho\utau^2 dP/dx$ is the equilibrium boundary layer parameter}
%  {\begin{tabular}{@{}cccc}\toprule
%   $\uinf$ ($\ms$) & $\dbl$ ($\milimeter$) & $m$  & $\beta_P$\\
%   \colrule
%   $20.3$ & $5.4$ & $0.024$ & $-0.022$\\
%   \botrule
%  \end{tabular}}
% \label{tab:TBLCarac}
% \end{table}
% EXTRAPOLATION TO THE WALL OF THE MEAN FLOW PROFILE
Using a procedure similar to the one described in \cite{Kendall2008}, we find the friction velocity $\utau$ by minimizing the mean deviation of the experimental points from an analytical profile for $y^+ < 140$.
We use the composite law of Monkewitz et al. \cite{Monkewitz2007} but we have verified that the use of other expressions, such as Musker's profile \cite[][]{Musker1979} leads to similar best-fit values of $\utau$. 
In figure \ref{fig:Velocity} we show the measured velocities (red dots) and the best-fitted profile (solid blue line). Values from the the best-fitted velocity profile then complement the measured points to estimate the values of the boundary layer parameters, reported in Table~\ref{tab:TBLCaracEstimate}.
The Reynolds number based on the displacement thickness $\Redstar=\uinf \dstar/\nu$ of the base flow is $\Redstar \approx 1000$ (while $\Retheta=\uinf \theta/\nu=750$). 
%The linear optimal growth results of \cite{Cossu2009} results obtained with the 
%According to the linear stability theory \cite[see][]{Cossu2009}, this value is large enough to see both inner and outer peaks of energy growth. 
\begin{table}
 \caption{Turbulent boundary layer base flow: parameters and variables estimated with the best-fit analytic profile. $C_f=2(\utau/\uinf)^2$ is the friction coefficient, $\Delta=\dstar \uinf/\utau$ is the Rotta-Clauser thickness and $H=\dstar/\theta$ denotes the shape factor.}
 {\begin{tabular}{@{}cccccccc}\toprule
  $\uinf$ ($\ms$) & $\utau$ ($\ms$) & $C_f\cdot10^3$ & $\dbl$ ($\milimeter$) & $\dstar$ ($\milimeter$) & $\theta$ ($\milimeter$) & $\Delta$ ($\milimeter$) & $H$  \\
  \colrule
  $20.3$ & $1.02$ & $5.05$ & $5.4$ & $0.79$ & $0.55$ & $15.7$ & $1.42$ \\
  \botrule
 \end{tabular}}
\label{tab:TBLCaracEstimate}
\end{table}

% \newpage
%%%%%%%%%%%%%%%%%%%%%%%%%%%%%%%%%%%%%%%%%%%%%%%%%%%%%%%%%%%%%%%%%%%%%%%%%%%%%%%%%%%%%%%%%%%%%%%%%%
\section{Forced large-scale coherent streaks} \label{sec:Streaks}

%-------------------------------------------------------------------------------------------------
\subsection{Parameters of interest} \label{sec:Configurations}

As already mentioned, roughness elements are used to generate nearly optimal streamwise vortices that induce nearly-optimal streamwise streaks. Following \cite{White2002} and \cite{Fransson2004} we use roughness elements of cylindrical cross-section which, because of their axi-symmetry, produce counter-rotating vortices of equal mean circulation even when immersed in irregular base flows. In the present study, the height $k$ of the elements is kept constant to $k=4\milimeter=0.8\dblo$, where $\dblo$ is the baseline boundary layer thickness at the position of the roughness elements.
Following previous studies \cite{Fransson2004,Fransson2005,Fransson2006} where nearly-optimal streaks were generated in the laminar boundary layer, we keep constant to $\lz/d=4$ the ratio of the spanwise spacing $\lz$ between the elements and their diameter $d$. This choice ensures that the vortices, and hence the streaks, have the same distribution of spanwise harmonic wavenumbers regardless of the chosen spanwise wavelength $\lz$. Several wavelength ranging from $3\dblo$ to $12\dblo$ are tested. These wavelengths roughly correspond to the most amplified wave-band obtained by computing the optimal energy growth of the base flow profile \cite{Cossu2009}.
The dimensions of the corresponding roughness elements are reported in Table~\ref{tab:ConfigFlatPlate}. For each configuration, the number of roughness elements is chosen so that the array spans $2/3$ of the test section width. The roughness elements array is always located at $x=x_0=110\milimeter$. 
\begin{table}
 \caption{Tested roughness elements configurations.}
 {\begin{tabular}{@{}cccccc}\toprule
 Config. & $\lz$ ($\milimeter$) & $d$ ($\milimeter$) & $\lz/d$ & $\lz/\dblo$ & $k/\dblo$ \\
 \colrule
 A & $15.8$ & $3.94$ & $4$ & $3$ & $0.8$ \\
 B & $26.8$ & $6.7$ & $4$ & $5$ & $0.8$ \\
 C & $33$ & $8.25$ & $4$ & $6$ & $0.8$ \\
 D & $40.$ & $10.$ & $4$ & $7.5$ & $0.8$ \\
 E & $50.8$ & $12.7$ & $4$ & $10$ & $0.8$ \\ 
 F & $65.6$ & $16.4$ & $4$ & $12$ & $0.8$ \\
 \botrule
 \end{tabular}}
\label{tab:ConfigFlatPlate}
\end{table}

%-------------------------------------------------------------------------------------------------
\subsection{The streamwise velocity field: large-scale coherent streaks} \label{sec:CoherentStreaks}

\begin{figure}
 \begin{center}
 \subfigure[]{
 \resizebox*{8cm}{!}{\includegraphics{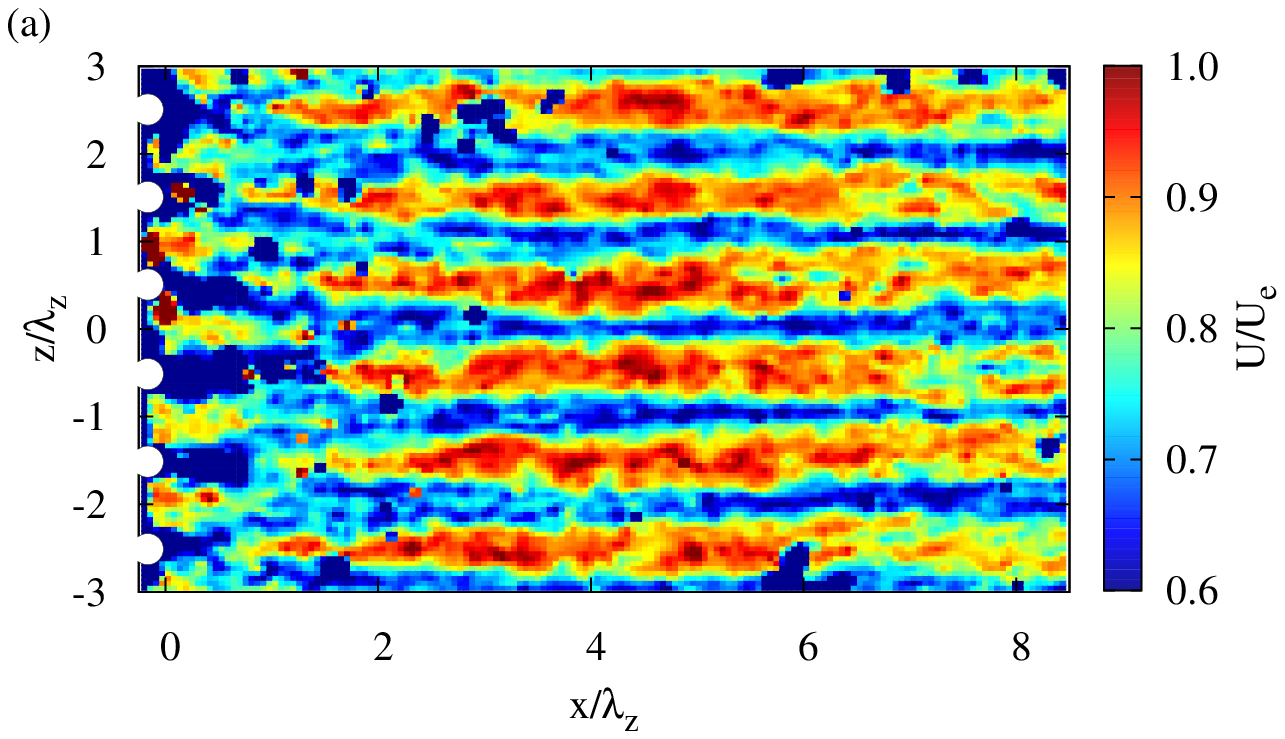}}
 }
 \subfigure[]{
 \resizebox*{8cm}{!}{\includegraphics{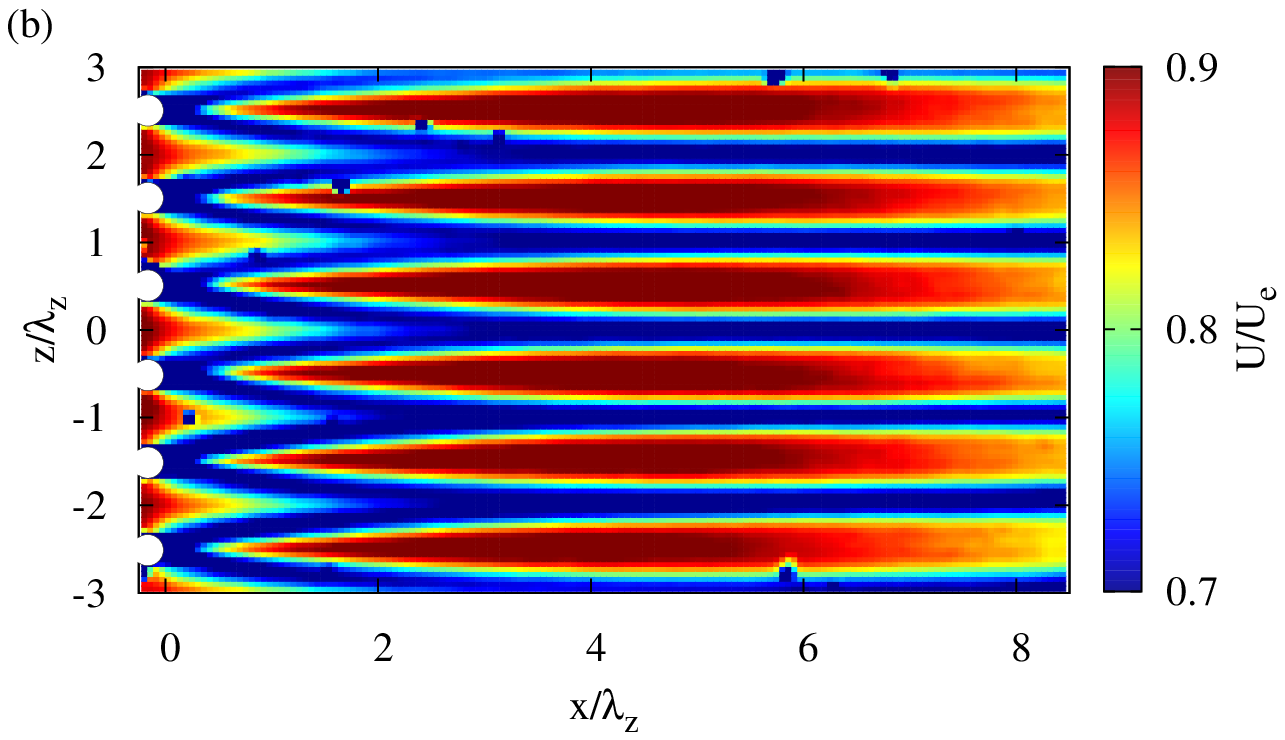}}
 }
 \caption{Visualization of large-scale coherent streaks downstream of the array of cylindrical roughness elements of configuration C. The velocity field displayed in the $x-z$ plane at half roughness height $Y=k/2$ from the wall. The flow is from left to right, the white disks on the left part of the figure indicate the position of the roughness elements. (a) Instantaneous streamwise velocity field. (b) Time-averaged streamwise velocity field $U/\uinf$.}
 \label{fig:CoherentStreaks}
 \end{center}
\end{figure}
\begin{figure}
 \begin{center}
 \resizebox*{8cm}{!}{\includegraphics{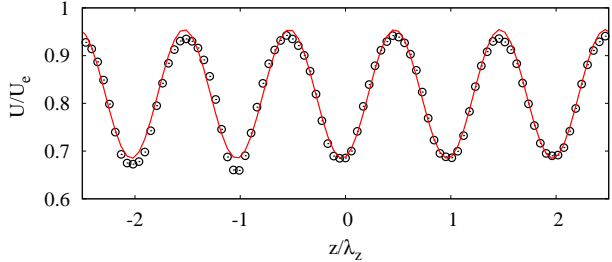}}
 \caption{Spanwise distribution of the mean velocity $U(x=4.\lz,Y=k/2,z/\lz)/\uinf$. \boldmath{$\circ$} PIV measured velocity distribution. {\textcolor{red}{\boldmath{$\relbar\!\relbar\!\relbar$}}} Best fit to the experimental data with a sinus function.}
 \label{fig:SinusStreaks}
 \end{center}
\end{figure}

We report in Figure~\ref{fig:CoherentStreaks} the streamwise velocity field measured in the plane at a distance of $Y=2mm=k/2$ from the wall for configuration C (see Table~\ref{tab:ConfigFlatPlate}). 
The streamwise velocity scaled on the free-stream velocity $\uinf$ is plotted against the streamwise and spanwise directions both scaled on the wavelength of the disturbance $\lz$ which, for this configuration, is equal to six times the boundary layer thickness. 
Large scale streaks are clearly visible with superposed small scale fluctuations in the instantaneous velocity field reported in Figure~\ref{fig:CoherentStreaks}(a). The large-scale streaks consist of an alternating pattern of high speed regions (hot contours) flanked with low speed streaks (cold contours).
The time-averaged velocity field (averaged over $600$ instantaneous fields) is reported in figure~\ref{fig:CoherentStreaks}(b). The averaging reveals the neat and well defined coherent  high speed and low speed streaks. Such a good definition of the large-scale streaks is due to the fact that the deterministic and steady forcing by the roughness elements induce  coherent streaks that are, in the present conditions, statistically steady, contrary to the unforced very large-scale coherent streaks. The measured coherent streaks are parallel to mean flow direction with the high speed streaks developing downstream of the roughness elements and the low speed streaks in between them, just like in the laminar boundary layer \cite{White2002,Fransson2004}. 
The spanwise distribution of the mean velocity $U(x=4.\lz,Y=k/2,z/\lz)/\uinf$ (open circles) is reported in Figure~\ref{fig:SinusStreaks}. The velocity modulation is  periodic in the spanwise direction with a spanwise wavelength equal to the spacing of the roughness elements in the array. For the considered moderate values of the streak amplitude the spanwise velocity profile is well fitted by a simple sinus function, displayed in the figure as the solid red line.  

%-------------------------------------------------------------------------------------------------
\subsection{Amplitude of the streaks} \label{sec:Amplitude}

We now come to the main objective of this study, which is to determine if coherent large-scale streaks developing in a turbulent mean flow can experience a transient growth of their amplitude as predicted by the recent theoretical analyses based on eddy viscosity models \cite{delAlamo2006,Pujals2009,Cossu2009,Hwang2009}. 
Like in the laminar case, in the present experimental framework the transient growth of the streaks is expected to happen in the streamwise direction. A measure of the amplitude of the streaks must be defined in order to measure this transient growth. 
%Linear stability theory predicts that large-scale turbulent streaks are only \emph{transiently} amplified through the lift-up effect. To confirm such behavior, we need a measure of the streaks' finite magnitude. 
Various definitions are available in the literature like the energy amplification used in linear stability analyses or the min-max criterion expressing the peak-to-peak difference between the velocities in the high and low speed streaks \cite[][]{Andersson2001}. %In the present study, the experimental apparatus does not permit to quantify the kinetic energy contained in the vortices created by the roughness elements nor to find the minimum velocity in low speed streaks.  
Here, following previous studies \cite{Hollands2009},  we estimate the amplitude of the streaks introducing a 'local' min-max criterion with: 
\begin{equation}
 \label{eq:MinMaxLocal}
 \amplstreaks\left(x,Y=k/2\right)=\frac{U(x,Y=k/2,z_{hsst})-U(x,Y=k/2,z_{lsst})}{2\uinf}
\end{equation}
where $z_{hsst}$ denotes the spanwise location of high speed streaks (i.e. $z/\lz=\pm0.5,\pm1.5,\pm2.5$) and $z_{lsst}$ denotes the location of the neighbouring low speed streaks (i.e. $z/\lz=0,\pm1,\pm2$). A more accurate value is then obtained performing a sliding averaging over the whole spanwise window. 
The so-defined amplitude maximize the velocity difference in the spanwise direction only for the given height $Y$ of the plane where the velocity is measured. This measure therefore represents only a lower bound on the amplitude defined by \cite{Andersson2001} where the min-max is found in both $z$ and $y$. This approximation has however already proven reasonable when compared to other similar definitions (see e.g. the definition by \cite{Hollands2009} applied to this configuration by \cite{Pujals2009b}) and is anyway sufficient to prove the existence of transient growth.

\begin{figure}
 \begin{center}
 \resizebox*{8cm}{!}{\includegraphics{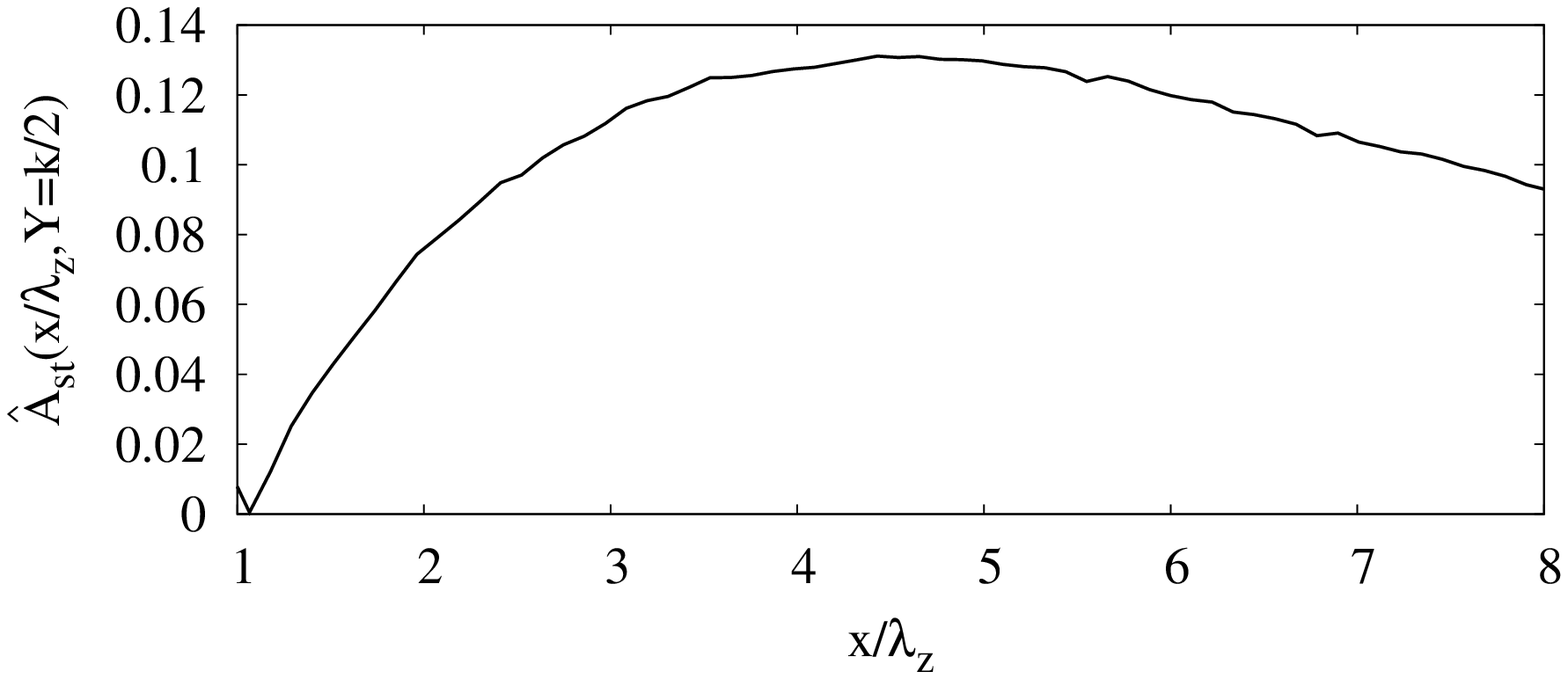}}
 \caption{Streamwise evolution of the finite amplitude $\amplstreaks\left(x/\lz,Y=k/2\right)$ of the streaks obtained with Configuration C.}
 \label{fig:TransientGrowth}
 \end{center}
\end{figure}

In figure~\ref{fig:TransientGrowth} we report the streamwise evolution $\amplstreaks\left(x,Y=k/2\right)$ of the amplitude of the coherent (averaged) streaks obtained with configuration C and displayed in figure~\ref{fig:CoherentStreaks}(a). This plot confirms what was already discernible on figure~\ref{fig:CoherentStreaks}(a):
The amplitude of the streaks grows in the downstream direction until a maximum value is reached, near $x-x_0 \approx 4 \lambda_z$ and then decays. The maximum amplitude is nearby $13\%$ of the free-stream velocity, a value of the order of the largest amplitude stable streaks experimentally forced in the laminar boundary layer \cite{Fransson2004,Fransson2005,Fransson2006}.
To the authors' knowledge, this is the first experimental observation of the transient growth of very-large-scale ($\lz=6\dblo$) coherent streaks developing in a turbulent mean flow and confirms without doubt the existence of a mean lift-up effect in turbulent shear flows.

%-------------------------------------------------------------------------------------------------
\subsection{Influence of the spanwise wavelength on the streaks amplitude} \label{sec:VaryingWavelength}

\begin{figure}
 \begin{center}
 \subfigure[]{
 \resizebox*{8cm}{!}{\includegraphics{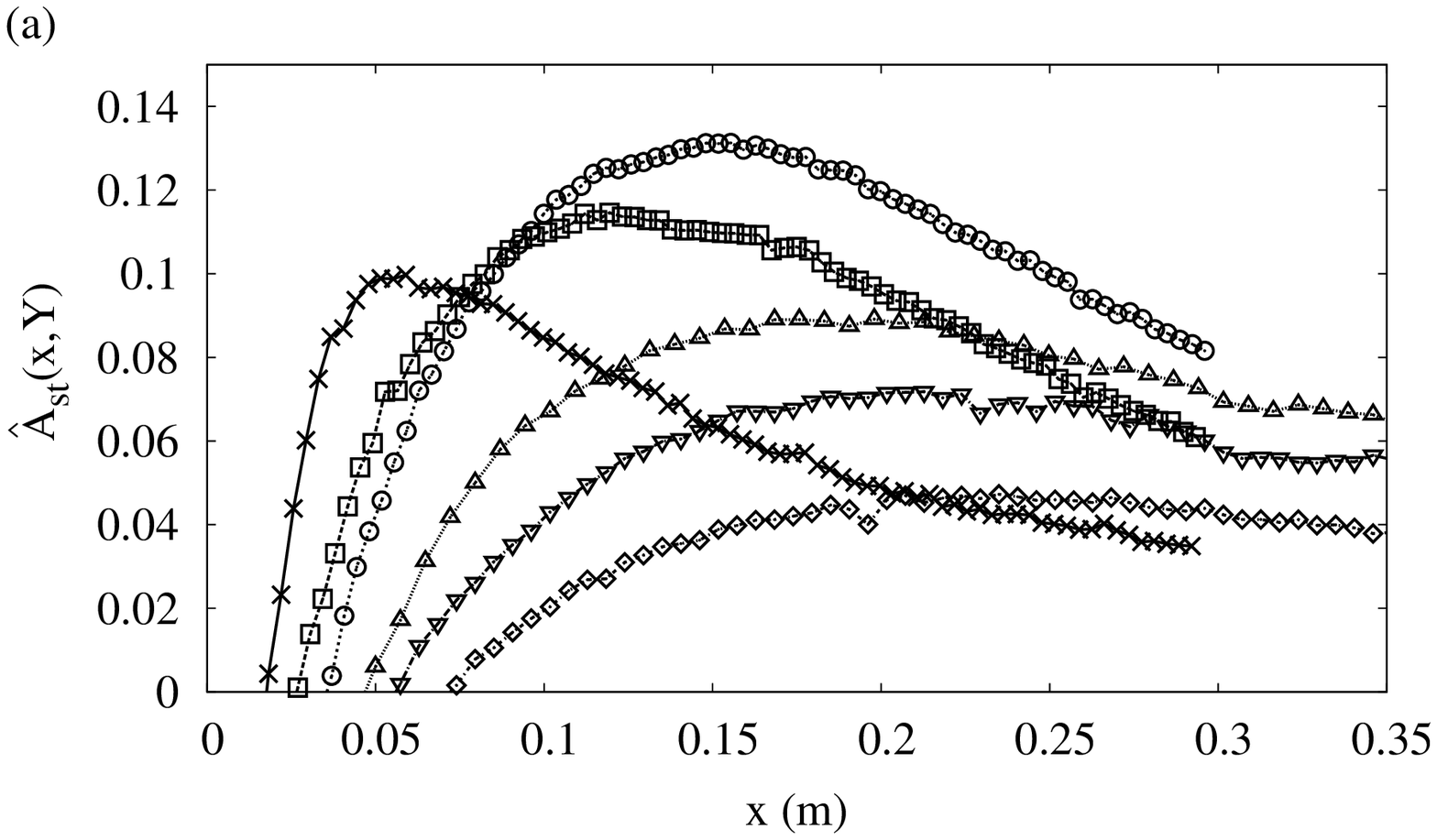}}
 }
 \subfigure[]{
 \resizebox*{8cm}{!}{\includegraphics{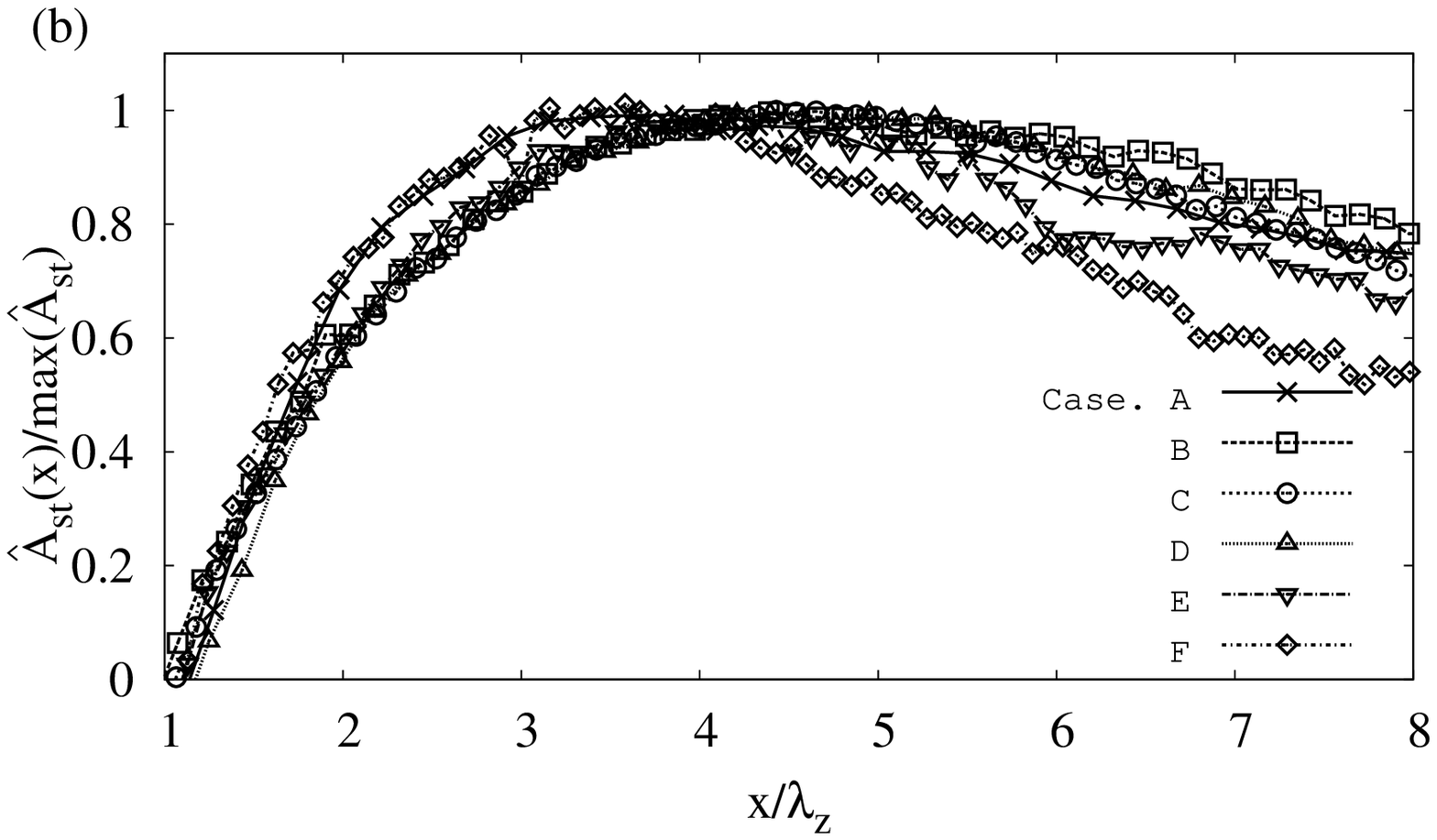}}
 }
 \caption{Finite amplitude of the streaks forced with all the configurations reported in Table~\ref{tab:ConfigFlatPlate}. (a) Amplitude plotted as a function of the distance from the roughness array $x$ expressed in meters. (b) Normalized amplitude plotted versus the distance from the roughness array scaled with the wavelength of the perturbation $x/\lz$.}
 \label{fig:AmplitudeComparison}
 \end{center}
\end{figure}

The influence of the spanwise wavelength on the streaks amplitude is analyzed by repeating the measures
for all the configurations described in Table~\ref{tab:ConfigFlatPlate}. For each wavelength from $\lz=3\dblo$ to $\lz=12\dblo$, large-scale coherent streaks similar to the ones presented in figure~\ref{fig:CoherentStreaks} are observed.
Their amplitudes are computed using always the same definition~\ref{eq:MinMaxLocal} and are reported
in figure~\ref{fig:AmplitudeComparison}.
In figure~\ref{fig:AmplitudeComparison}(a) the amplitudes are plotted versus the streamwise coordinate $x$.  
For all the considered spanwise wavelengths transient growth of the streak amplitude is observed. 
The maximum amplitude and the streamwise station where the maximum is reached depends on the wavelength of the disturbance. 
In figure~\ref{fig:AmplitudeComparison}(b) the same data are replotted, with the maximum amplitudes normalized to unity and the streamwise coordinate rescaled with the wavelength $\lz$ of the perturbation. All the curves collapse reasonably well, except possibly the extreme cases A and F. For all the considered spanwise wavelengths the artificially forced coherent streaks reach their maximum amplitude at $x_{max}=4\sim5\lz$ downstream of the roughness array. 
The proportionality of $x_{max}$ to $\lz$ is reminiscent of the proportionality of $t_{max}$ on $\lz$ observed in the temporal optimal growth analyses \cite[see][]{delAlamo2006,Pujals2009,Cossu2009} and of the proportionality of the lifetime of large-scale structures to their size \cite[see][]{delAlamo2006b}. 
% PLOT WITH AMAX VERSUS LAMBDAZ AND COMPARISON WITH THE THEORETICAL PREDICTIONS

\begin{figure}
 \begin{center}
 \resizebox*{8cm}{!}{\includegraphics{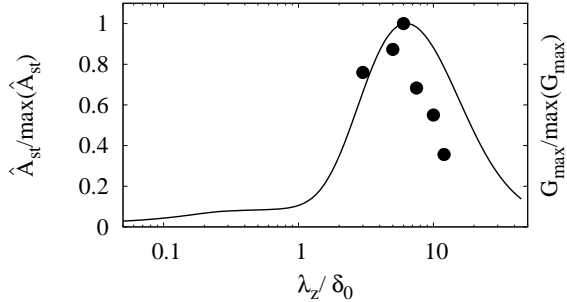}}
 \caption{Black discs: experimentally measured maximum streak amplitude $\amplstreaks^{max}$ scaled with the global maximum amplitude $\max(\amplstreaks^{max})$ versus the spanwise wavelenght $\lz$ scaled with the boundary layer thickness $\dblo$. Solid line: Optimal growth $G_{max}$ data from \cite{Cossu2009} normalized on the \emph{global} maximum and computed for streamwise uniform perturbations and the same mean base flow.}
 \label{fig:ExpVsLin}
 \end{center}
\end{figure}

The dependence of the maximum streak amplitude on $\lz$ is reported in figure~\ref{fig:ExpVsLin}. In that figure, the maximum amplitude $\amplstreaks^{max}$ obtained for each wavelength is scaled with the \emph{global} maximum amplitude $\max(\amplstreaks^{max})$ and reported as black disks. A maximum of the amplitude is observed for $\lz \approx 6\dblo$ (case C). This value is well in the range of the theoretical predictions \cite{Cossu2009}. We therefore report with a solid line in figure~\ref{fig:ExpVsLin} the optimal growth $G_{max}$ data computed in \cite{Cossu2009} and normalized with the \emph{global} maximum. We remind that these optimal growth data have been obtained for streamwise uniform perturbations for the turbulent mean flow profile of \cite{Monkewitz2007} at the same Reynolds number $\Redstar \approx 1000$ of the experimental base flow. The optimal growth data from \cite{Cossu2009} and the experimental data agree reasonably well eventhough the two compared quantities are different: $\amplstreaks$ is an amplitude while $G_{max}$ stands for an energy amplification. The implicit assumption made is that the initial amplitude of the vortices generated by the roughness elements does not change much with $\lz$. The important point is however that both the optimal growth analysis and the experimental data indicate that the maximum amplification of the coherent streaks by the mean lift-up effect is obtained for spanwise wavelengths of $\lz \approx 6\dblo$ at this Reynolds number and that neighbouring very large-scale wavelenghts $\lz$ are also noticeably amplified.
%This result confirms that a selection mechanism is operated by the base flow. In the present study, this value is between $5\%$ and $13\%$ of the free stream velocity $\uinf$. In addition, the distance over which the maximum amplitude is attained increases with the wavelength. This later result is also in agreement with the linear stability theory which predicts that the larger the wavelength is, the longer is the time of optimal growth. 

% \newpage
%%%%%%%%%%%%%%%%%%%%%%%%%%%%%%%%%%%%%%%%%%%%%%%%%%%%%%%%%%%%%%%%%%%%%%%%%%%%%%%%%%%%%%%%%%%%%%%%%%
\section{Conclusions and outlook} \label{sec:Conclusions}

Recent theoretical analyses have predicted the amplification of coherent streaks via a mean lift-up effect in turbulent shear flows. In particular, for the zero-pressure gradient turbulent boundary layer, it was found \cite{Cossu2009}  that the most amplified structures are large-scale streamwise uniform coherent streaks with spanwise wavelength of $6-8 \dbl$. Naturally occurring large-scale streaks with such a large spanwise wavelenght have not been experimentally observed in turbulent boundary layers. Even the very existence of a transient growth of turbulent coherent streaks at very large-scale could be questioned as this growth has never been experimentally observed. 
The twofold goal of this study was therefore to show if transient growth of turbulent coherent streaks could be experimentally observed and, if yes, to determine the amplified wave-band and compare it to the predictions of optimal growth theory.

The main idea underlying our study is that in order to have reproducible results, at least in the mean, one has to artificially force the coherent streaks. We have therefore used arrays of small cylindrical roughness elements to introduce streamwise nearly-optimal coherent vortices in the turbulent boundary layer at a Reynolds number of $\Redstar=1000$. These vortices are expected to induce the growth of coherent streaks. The main results are as follows: 
\begin{enumerate}
\item Very large-scale coherent streaks can be artificially forced in the turbulent boundary layer. These coherent streaks are in the mean oriented as the mean flow and stable for the considered cases.
\item The artificially forced large-scale coherent streaks experience a transient growth in space, similarly to the growth observed for streaks forced in the same way in a laminar boundary layer. They can reach similar finite moderate amplitudes $\approx 13\% \uinf$.
\item All the amplitude curves are observed to collapse on a single one when rescaled on the maximum amplitude and the spanwise wavelength of the perturbation. The maximum amplitude of the streaks is attained for $x_{max}=4\sim5\lz$ regardless of the wavelength.
\item In agreement with results from the linear optimal perturbations theory, for this Reynolds number, the most amplified wavelength is  $\approx 6\dblo$. 
\end{enumerate}

These results confirm the existence of a mean lift-up effect predicted by the recent theoretical approaches based on strong simplifying assumptions such as the modelling of the Reynolds stresses with an eddy viscosity, the locally parallel flow assumption  and the use of the temporal theory. 
Given these strong assumptions it is even more surprising to find that there is also a very reasonable quantitative agreement between experiments and theory concerning the most amplified spanwise wavelength. We here confirm that the most amplified spanwise wavelengths are $\approx 6\dblo$ at $\Redstar \approx 1000$.
%These results are important because they show that not only near wall structures, but also very large-scale coherent structures are able to extract energy from the mean flow via the described mean lift-up. 
Of course, experiments at larger Reynolds numbers are necessary to confirm the present findings and rule out any possible low Reynolds number effect.

%Very large-scale vortices with $\lz\in\left[3\dblo,12\dblo\right]$ are able to generate very large-scale coherent turbulent streaks further downstream. These streaks are amplified through the lift-up effect and modify the base flow at leading order in spite of non-linear mechanisms and 'real' turbulence (we recall that linear stability theory 'simply' relies on an eddy viscosity profile to describe the turbulence dynamics). 
% DISCUSS RELEVANCE OF THE RESULTS
%These results tend to prove that lift-up effect can be used to amplify large-scale perturbations in a wall-bounded turbulent flow. A still open question is the influence of the amplitude of the streaks on their behavior. In laminar flow cases, large amplitude streaks are known to undergo a secondary instability which lead to breakdown and eventually transition to turbulence. In the turbulent case, similar behaviour has not been observed in the present experiments. One reason for this could be the amplitudes attained here that are too small to ignite such kind of secondary instability. If one could trigger it, the resulting dynamics would be of great interest for a better understanding of streaks/vortices generation mechanisms. 
% In the core region of turbulent Couette flow, \cite{Kitoh2008} also confirmed that very large-scale streaks can be artificially forced.
% DISCUSS IMPLICATIONS FOR CONTROL
Furthermore, these results open the way to the use of the lift-up of coherent streaks for control purposes. Such a large-scale control is promising because it has been recently acknowledged that natural large-scale structures influence the near-wall dynamics \cite[][]{Hutchins2007b} and in particular the development of near-wall streaks \cite[][]{Itano2001}. Such interaction between forced large-scale and small-scale structures have been used to reduce skin-friction drag \cite[see][]{Schoppa1998,Iuso2002}. These results suggest that similarily very large-scale coherent streaks such as those forced in the present study can have a strong influence on the near wall streaks dynamics. 

In the laminar Blasius boundary layer, nearly optimal streaks were successfully used to stabilize Tollmien-Schlichting waves and then delay transition to turbulence \cite[][]{Fransson2005,Fransson2006}. Due to their strong amplification, the forced streaks modify the mean velocity profiles at leading order. The spanwise averaged shape factor of the resulting velocity profile is lower than the unforced case implying a fuller, and then more resistant, mean velocity profile. In the present turbulent case, similar modification of the mean velocity profiles is to be expected since we have observed a strong modulation of the mean velocity field in presence of the streaks. Such modification could result in a turbulent mean velocity profile with a lower shape factor. Current active research show that the large amplification of the coherent turbulent streaks could indeed be beneficial for separation control \cite{Pujals2010a}.

\bibliographystyle{osa}
%\bibliography{carlo}
\bibliography{greg}

\label{lastpage}

\end{document}